\documentclass[11pt]{article}

\usepackage{fullpage,amsmath,amsthm}
\usepackage{amssymb,latexsym}

\theoremstyle{plain}
\newtheorem{theorem}{Theorem}

\newtheorem{lemma}[theorem]{Lemma}
\theoremstyle{definition}

\def\01{\{0,1\}}

\newcommand{\Exp}{{\mathbb{E}}}
\newcommand{\alphapm}{\textsc{$\alpha$PM}}
\newcommand{\eps}{\varepsilon}
\newcommand{\ket}[1]{|#1\rangle}

\newcommand{\ketbra}[2]{|#1\rangle\langle#2|}
\newcommand{\inprod}[2]{\langle{#1},{#2}\rangle} 
\newcommand{\norm}[1]{\mbox{$\parallel{#1}\parallel$}}

\newcommand{\COMMENT}[1]{}

\begin{document}

\title{Exponential separation for one-way quantum communication complexity, with applications to cryptography}
\author{
Dmitry Gavinsky\thanks{Supported in part by Canada's NSERC.}\\IQC, University of Waterloo
 \and Julia Kempe\thanks{Part of this work was done while this author was with LRI \& CNRS, Univ.~de Paris-Sud, Orsay, France.
 Supported in part by ACI S\'ecurit\'e
Informatique SI/03 511 and ANR AlgoQP grants of the French Research Ministry, by an Alon Fellowship of the
Israeli Higher Council of Academic Research, by a grant of the Israeli Science Foundation and by the European
Commission under the Integrated Project Qubit Applications (QAP) funded by the IST directorate as Contract
Number 015848.}\\School of Computer Science\\ Tel Aviv University
 \and
Iordanis Kerenidis$^\dagger$\\
CNRS \& LRI\\
Univ.~de Paris-Sud, Orsay
 \and Ran Raz\thanks{Part of this work was done when this author
visited Microsoft Research, Redmond, supported by ISF and BSF grants.}\\
Faculty of Mathematics\\
Weizmann Institute
 \and
Ronald de Wolf\thanks{Supported by a Veni grant from the Netherlands Organization for Scientific Research
(NWO) and also partially supported by the European Commission under the Integrated Project Qubit
Applications (QAP) funded by the IST directorate as Contract Number 015848.}\\CWI\\ Amsterdam
 }
\date{}
\maketitle

\begin{abstract}
We give an exponential separation between one-way quantum and classical communication protocols for a
partial Boolean function (a variant of the Boolean Hidden Matching Problem of Bar-Yossef et al.)
Earlier such an exponential separation was known only for a relational problem.
The communication problem corresponds to a \emph{strong extractor} that fails against a small
amount of \emph{quantum} information about its random source.
Our proof uses the Fourier coefficients inequality of Kahn, Kalai, and Linial.

We also give a number of applications of this separation. In particular, we show that there are privacy
amplification schemes that are secure against classical adversaries but not against quantum adversaries;
and we give the first example of a key-expansion scheme in the model of bounded-storage cryptography
that is secure against classical memory-bounded adversaries but not against quantum ones.
\end{abstract}

\newpage

\section{Introduction}

One of the main goals of quantum computing is to exhibit
problems where quantum computers are much faster (or otherwise better) than classical computers. Preferably
exponentially better. The most famous example, Shor's efficient quantum factoring algorithm~\cite{shor:factoring}, constitutes a
separation only if one is willing to believe that efficient factoring is impossible on a classical
computer---proving this would, of course, imply P$\neq$NP. One of the few areas where one can establish
\emph{unconditional} exponential separations is communication complexity.

Communication complexity is a central model of computation, first defined by Yao~\cite{yao:distributive}, that has found applications in many areas~\cite{kushilevitz&nisan:cc}. In this model,
two parties, Alice with input $x$ and Bob with input $y$, collaborate to solve some computational problem
that depends on both $x$ and $y$. Their goal is to do this with minimal communication. The problem to be solved
could be a function $f(x,y)$ or some relational problem where for each $x$ and $y$, several outputs
are valid. The protocols could be \emph{interactive (two-way)}, in which case Alice and Bob take turns sending messages to each other; \emph{one-way}, in which case Alice sends a single message to Bob
who then determines the output; or \emph{simultaneous}, where Alice and Bob each pass one message to a
third party (the \emph{referee}) who determines the output.  The bounded-error \emph{communication complexity} of the
problem is the worst-case communication of the best protocol that gives (for every input $x$ and $y$)
a correct output with probability at least $1-\eps$, for some fixed constant $\eps\in[0,1/2)$, usually $\eps=1/3$.

Allowing the players to use \emph{quantum} resources can reduce the communication complexity
significantly. Examples of problems where quantum communication gives exponential savings were
given by Buhrman, Cleve, and Wigderson for one-way and interactive protocols with zero error
probability~\cite{BuhrmanCleveWigderson98}; by Raz for bounded-error interactive protocols~\cite{raz:qcc};
and by Buhrman, Cleve, Watrous, and de Wolf for bounded-error simultaneous protocols~\cite{bcww:fp}.
The first two problems are partial Boolean functions, while the third one is a total Boolean function.
However, the latter separation does not hold in the presence of public coins.\footnote{In fact, whether there exists a superpolynomial separation for a total Boolean function in the presence of public coins is one of the main open questions in the area of quantum communication complexity.}
Bar-Yossef, Jayram, and Kerenidis~\cite{bjk:q1way} showed an exponential separation for one-way protocols
and simultaneous protocols with public coins, but they only achieved this for a relational
problem, called the \emph{Hidden Matching Problem} (HMP). This problem can be solved efficiently by one
quantum message of $\log n$ qubits, while classical one-way protocols need to send nearly $\sqrt{n}$ bits to
solve it. Nevertheless, Boolean functions are much more natural objects than relations both in the model of
communication complexity and in the cryptographic settings that we consider later in this paper.
Bar-Yossef et al.\ stated a Boolean version of their problem (a partial Boolean function)
and conjectured that the same quantum-classical gap holds for
this problem as well.

\subsection{Exponential separation for a variant of Boolean Hidden Matching}

In this paper we prove an exponential quantum-classical one-way communication gap
for a variant of the Boolean Hidden Matching Problem of~\cite{bjk:q1way}.
Let us first state a non-Boolean communication problem.
Suppose Alice has an $n$-bit string $x$, and Bob has a sequence $M$ of $\alpha n$ disjoint
pairs $(i_1,j_1),(i_2,j_2),\ldots,(i_{\alpha n},j_{\alpha n})\in[n]^2$, for some parameter $\alpha\in(0,1/2]$.
This $M$ may be viewed as a partial matching on the graph whose vertices are the $n$ bits $x_1,\ldots,x_n$.
We call this an \emph{$\alpha$-matching}.
Together, $x$ and $M$ induce an $\alpha n$-bit string $z$ defined by the parities of the $\alpha n$ edges:
$$
z=z(x,M)=(x_{i_1}\oplus x_{j_1}),(x_{i_2}\oplus x_{j_2}),\ldots,(x_{i_{\alpha n}}\oplus x_{j_{\alpha n}}).
$$
Suppose Bob wants to learn some information about $z$.
Let $x\in\01^n$ be uniformly distributed,
and $M$ be uniform over the set ${\cal M}_{\alpha n}$ of all $\alpha$-matchings.
Note that for any fixed $M$, a uniform distribution on $x$ induces a uniform distribution on $z$.
Hence Bob (knowing $M$ but not $x$) knows nothing about $z$: from his perspective it is uniformly distributed.
But now suppose Alice can send Bob a short message. How much can Bob learn about $z$, given that message and $M$?

The situation is very different depending on whether the message is quantum or classical.
Modifying the protocol of~\cite{bjk:q1way}, it is easy to show that a short \emph{quantum} message of about $\log(n)/2\alpha$
qubits allows Bob to learn a bit at a random position in the string $z$. This already puts a lower bound of one
on the total variation distance between Bob's distribution on $z$ and the uniform $\alpha n$-bit distribution.

What about a short \emph{classical} message? Using the Birthday Paradox, one can show that if Alice sends
Bob about $\sqrt{n/\alpha}$ bits of $x$, then with constant probability there will be one edge
$(i_\ell,j_\ell)$ for which Bob receives both bits $x_{i_\ell}$ and $x_{j_\ell}$.
Since $z_\ell=x_{i_\ell}\oplus x_{j_\ell}$, this gives Bob a bit of information about $z$.
Our key theorem says that this classical upper bound is essentially optimal:
if Alice sends much fewer bits, then from Bob's perspective the string $z$ will be
close to uniformly distributed, so he does not even know one bit of $z$.

In order to be able to state this precisely, suppose Alice is deterministic and sends $c$ bits of communication.
Then her message partitions the set of $2^n$ $x$'s into $2^c$ sets, one for each message.  
A typical message will correspond to a set $A$ of about $2^{n-c}$ $x$'s.  Given this message,
Bob knows the random variable $X$ is drawn uniformly from this set $A$ and he knows $M$, which is his input.
Hence his knowledge of the random variable $Z=z(X,M)$ is fully described by the distribution
$$
p_M(z)=\Pr[Z=z\mid\mbox{given $M$ and Alice's message}]=\frac{|\{x\in A\mid z(x,M)=z\}|}{|A|}.
$$
Our main technical result says that if the communication $c$ is much less than $\sqrt{n/\alpha}$ bits,
then for a typical message and averaged over all matchings $M$, this distribution is very close to uniform 
in total variation distance.
In other words: most of the time, Bob knows essentially nothing about~$z$. 

\begin{theorem}\label{thseparationforalphapm}
Let $x$ be uniformly distributed over a set $A\subseteq\01^n$ of size $|A|\geq 2^{n-c}$ for some $c\geq 1$, 
and let $M$ be uniformly distributed over the set ${\cal M}_{\alpha n}$ of all $\alpha$-matchings, 
for some $\alpha\in(0,1/4]$.
There exists a universal constant $\gamma>0$ (independent of $n$, $c$, and $\alpha$), such that for all $\eps>0$:
if $c\leq \gamma\eps\sqrt{n/\alpha}$ then
$$
\Exp_M\left[\norm{p_M - U}_{tvd}\right]\leq\eps.
$$
\end{theorem}

We prove Theorem~\ref{thseparationforalphapm} using the Fourier coefficients 
inequality of Kahn, Kalai, and Linial~\cite{kkl:influence},
which is a special case of the Bonami-Beckner inequality~\cite{Bon70,Bec75}. 
We remark that Fourier analysis has been previously used in communication complexity 
by Raz~\cite{raz:ccfourier} and Klauck~\cite{klauck:qcclower}.

This result allows us to turn the above communication problem into a
partial Boolean function, as follows. Again we give Alice input $x\in\01^n$, while Bob now receives two
inputs: a partial matching $M$ as before, and an $\alpha n$-bit string $w$.  The promise on the input is that
$w$ is either equal to $z=z(x,M)$, or to its complement $\overline{z}$ (i.e.~$z$ with all bits flipped). 
The goal is to find out which of these two possibilities is the case. We call this communication problem $\alphapm$, 
for ``$\alpha$-Partial Matching''. As mentioned before, Alice can allow Bob 
to learn a random bit of $z$ with high probability by sending him 
an $O(\log(n)/\alpha)$-qubit message. Knowing one bit $z_\ell$ of $z$ suffices
to compute the Boolean function: just compare $z_\ell$ with $w_\ell$.  In contrast, if Alice sends Bob much
less than $\sqrt{n/\alpha}$ classical bits, then Bob still knows essentially nothing about $z$. In
particular, he cannot decide whether $w=z$ or $w=\overline{z}$! This gives the following separation result
for the classical and quantum one-way communication complexities (with error probability fixed to $1/3$, say):

\begin{theorem}\label{thccseparation}
Let $\alpha\in(0,1/4]$.
The classical bounded-error one-way communication complexity of the $\alpha$-Partial Matching problem is
$R^1({\alphapm})=\Theta(\sqrt{n/\alpha})$, while the quantum bounded-error
one-way complexity is $Q^1({\alphapm})=O(\log(n)/\alpha)$
\end{theorem}

Fixing $\alpha$ to $1/4$, we obtain the promised exponential quantum-classical separation for 
one-way communication complexity of $O(\log n)$ qubits vs $\Omega(\sqrt{n})$ classical bits.

\medskip

{\bf Remarks.}
The earlier conference version of this paper~\cite{gkkrw:1way}
had two different communication problems, establishing an exponential one-way separation for both 
of them in quite different ways. The present paper unifies these two approaches to something substantially simpler.

The original Boolean Hidden Matching Problem stated in~\cite{bjk:q1way} is our \alphapm\ with $\alpha=1/2$
(i.e.~$M$ is a perfect matching). 
Theorem~\ref{thccseparation}, on the other hand, assumes $\alpha\leq 1/4$
for technical reasons. By doing the analysis in Section~\ref{secmainproof} a bit more carefully, 
we can prove Theorem~\ref{thccseparation} for every $\alpha$ that is bounded away from 1/2.
Note that if $\alpha=1/2$, then the parity of $z=z(x,M)$ equals the parity of $x$, 
so by communicating the parity of $x$ in one bit, Alice can give Bob one bit of information about $z$. 
The conference version of this paper showed that one can
prove a separation for the case where $M$ is a perfect matching if the promise is that $w$ is
``close'' to $z$ or its complement (instead of being \emph{equal} to $z$ or its complement). One can think of
$w$ in this case as a ``noisy'' version of $z=z(x,M)$ (or its complement), while the $w$ of our current version
can be thought of as starting from a perfect matching $M'$, and then ``erasing'' some of the $n$ bits of 
the string $z(x,M')$ to get the $\alpha n$-bit string $z$ (or its complement).

The separation given here can be modified to a separation in the simultaneous message
passing model, between the models of classical communication with shared entanglement and classical
communication with shared randomness. Earlier, such a separation was known only for a relational
problem~\cite{bjk:q1way,gkrw:identification}, not for a Boolean function.

\subsection{Application: privacy amplification}\label{ssecprivampl}

Randomness \emph{extractors} extract almost uniform randomness from an \emph{imperfect} (i.e.\
non-uniform) source of randomness $X$ with the help of an independent uniform seed $Y$.
With a bit of extra work (see Section~\ref{secextractor}), Theorem~\ref{thseparationforalphapm} actually
implies that our function $z:\01^n\times{\cal M}_{\alpha n}\rightarrow\01^{\alpha n}$ is an extractor:
\begin{quote}
If $X\in\01^n$ is a random variable with min-entropy at least $n-\gamma\eps\sqrt{n/\alpha}$ (i.e.~$\max_x
\Pr[X = x]\leq 2^{-(n-\gamma\eps\sqrt{n/\alpha})}$) and $Y$ is a random variable uniformly distributed over
${\cal M}_{\alpha n}$, then the random variable $Z:=z(X,Y)$ is $\eps$-close to the uniform distribution on
$\01^{\alpha n}$.
\end{quote}
It is in fact a \emph{strong} extractor: the pair $(Y,Z)$ is $\eps$-close to the
uniform distribution on ${\cal M}_{\alpha n}\times\01^{\alpha n}$.%
\footnote{Note that $\Exp_M\left[\norm{p_M - U}_{tvd}\right]=\norm{(Y,Z)-U}_{tvd}$, where `$U$'
on left and right is uniform over different domains.}
Informally, this says that if there is a lot of uncertainty about $X$, then $Z$ will be close to uniform
even if $Y$ is known.\footnote{It should be noted that the parameters of our extractor are quite bad,
as far as these things go.  First, the uniform input seed $Y$ takes about $\alpha n\log n$ bits to describe,
which is more than the $\alpha n$ bits that the extractor outputs; in a good extractor, we want the seed length
to be much shorter than the output length.  Second, our assumed lower bound on the initial min-entropy
is quite stringent.  Finally, the distance from uniform can be made polynomially small in $n$
(by putting an $n-n^{1/2-\eta}$ lower bound on the min-entropy of $X$) but not exponentially small, which is
definitely a drawback in cryptographic contexts. Still, this extractor suffices for our purposes here.}

Extractors have found numerous applications in computer science, in particular in complexity theory (see
e.g.~\cite{shaltiel:extractor} and the references therein) and cryptography. One important cryptographic
application is that of \emph{privacy amplification}, which was introduced in
\cite{Brassard:privacyamplification,ImpagliazzoLL:privacy}. In this setting, Alice and Bob start with a
shared random variable $X$ about which the adversary has some partial information $m(X)$ and their goal is to
generate a secret key $Z$ about which the adversary has very little information. They can achieve this by
communicating an independent uniform seed $Y$ over a \emph{public} authenticated channel, and using a strong
extractor to generate the key $Z(X,Y)$. Using the extractor we define here, the resulting $\alpha n$-bit key
$Z=z(X,Y)$ is $\eps$-close to uniform if the adversary's view of $X$ has min-entropy at least
$n-\gamma\eps\sqrt{n/\alpha}$. Thus, assuming a certain upper bound on the number of bits of $m(X)$, the key
$Z$ is $\eps$-secure despite the fact that the adversary can learn $Y$ completely by tapping the public
channel. Notice, however, that this classically-secure privacy amplification scheme is insecure against a
\emph{quantum} adversary: if the adversary stores a uniform superposition of the bits of $x$, then when later
$Y$ is revealed, she can learn a random bit of $Z$ with good probability. Thus we have an example of a
privacy amplification scheme that is secure against classical adversaries with $o(\sqrt{n})$ bits of storage,
but insecure against quantum adversaries with much less quantum storage.

This dependence of the security on whether the adversary has quantum or classical memory is quite surprising,
particularly in light of the following two facts. First, privacy amplification based on two-universal hashing
provides exactly the same security against classical and quantum adversaries. The length of the key that can
be extracted is given by the min-entropy both in the classical
(\cite{Brassard:privacyamplification,ImpagliazzoLL:privacy}) and the quantum case
(\cite{kmr:powerofqmem,renner&koenig:privampl}, \cite[Ch.~5]{renner:thesis}). Second, K\"onig and
Terhal~\cite{KoenigTerhal06} have recently shown that for protocols that extract just \emph{one} bit, the
level of security against a classical and a quantum adversary (with the same information bound) is comparable.

\subsection{Application: key-expansion in the bounded-storage model}

In privacy amplification, we can ensure that the adversary has much uncertainty about the random variable $X$ by assuming that he has only bounded storage. The idea of basing cryptography on
storage-limitations of the adversary was introduced by Maurer~\cite{Maurer92b}
with the aim of implementing information-theoretically secure \emph{key-expansion}.
In this setting, a large random variable $X$ is publicly but only temporarily available. Alice
and Bob use a shared secret key $Y$ to extract an additional key $Z=Z(X,Y)$ from $X$, in such a way
that the adversary has only limited information about the pair $(Y,Z)$.
``Limited information'' means that the distribution
on $(Y,Z)$ is $\eps$-close to uniform even when conditioned on the information about $X$ that the
adversary stored. Thus Alice and Bob have expanded their shared secret key from $Y$ to $(Y,Z)$.
Aumann, Ding, and Rabin~\cite{AumannDR:boundedstorage} were the first to prove a bounded-storage scheme
secure, and essentially tight constructions have subsequently been
found~\cite{DzieMaurer:boundedstorage,Lu:boundedstorage,Vadhan:boundedstorage}.

It is an important open question whether any of these constructions remain secure if the adversary is allowed
to store \emph{quantum} information. One may even conjecture that a bounded-storage protocol secure against
classical adversaries with a certain amount of memory, should be roughly as secure against \emph{quantum}
adversaries with roughly the same memory bound. After all, Holevo's theorem~\cite{holevo} tells us that $k$
qubits cannot contain more information than $k$ classical bits. However, a key-expansion scheme based on our
extractor refutes this conjecture. The scheme is essentially the same as the above privacy amplification
scheme: Alice and Bob will compute $Z:=z(X,Y)$ by applying our extractor to $X$ and $Y$. If the adversary's
memory is bounded by $\gamma\eps\sqrt{n/\alpha}$ bits then $Z$ will be $\eps$-close to uniform from the
adversary's perspective.  On the other hand, $O(\log n)$ qubits of storage suffice to learn one or more bits
of information about $Z$, given $Y$, which shows that $(Y,Z)$ is not good as a key against a quantum
adversary. Thus we have an example of a key-expansion scheme that is secure against classical adversaries
with $o(\sqrt{n})$ bits of storage, but insecure against quantum adversaries  even with exponentially less
quantum storage.

\subsection{Application: a separation in the streaming model}

In the \emph{streaming model} of computation, the input is given as a stream of bits and the algorithm is supposed
to compute or approximate some function of the input, having only space of size $S$ available. See for
instance~\cite{ams:frequency,muthukrishnan:datastreams}.
There is a well-established connection between one-way communication complexity and the streaming model:
if we view the input as consisting of two consecutive parts $x$ and $y$, then the content of the memory
after $x$ has been processed, together with $y$, contains enough information to compute $f(x,y)$. Hence, a space-$S$ streaming algorithm for $f$ implies a one-way protocol for $f$ of communication $S$ with the same success probability.
The classical lower bound for our Boolean communication complexity problem, together with the observation that our quantum protocol can be implemented in the streaming model, implies a separation between the quantum and classical streaming model. Namely, there is a partial Boolean function $f$ that can be computed in the streaming model with small error probability using quantum space of $O(\log n)$ qubits, but requires $\Omega(\sqrt{n})$ bits if the space is classical.

Le Gall~\cite{legall:onlinespace} constructed a problem that can be solved in the streaming model using
$O(\log n)$ qubits of space, while any classical algorithm needs $\Omega(n^{1/3})$ classical bits. His $\log
n$-vs-$n^{1/3}$ separation is a bit smaller than our $\log n$-vs-$\sqrt{n}$, but his separation is for a
\emph{total} Boolean function while ours is only partial (i.e.\ requires some promise on the input). Le
Gall's result predates ours, though we only learned about it after finishing the conference version of our
paper. We remark also that Le Gall's separation holds only in the streaming model variant where the bits
arrive in order, while ours holds in the more general model where we allow the different pieces of the input
to arrive in any order. The algorithm starts out with a $\log n$-qubit superposition
$\frac{1}{\sqrt{n}}\sum_{i=1}^n\ket{i}$. Whenever a bit $x_i$ streams by in the input, the algorithm applies
a unitary transformation that maps $\ket{i}\mapsto (-1)^{x_i}\ket{i}$. 
Whenever an edge $(i_\ell,j_\ell)$ streams by, the
algorithm measures with operators $E_1=\ketbra{i_\ell}{i_\ell}+\ketbra{j_\ell}{j_\ell}$ and $E_0=I-E_1$.  And
whenever a bit $(i_\ell,j_\ell,w_\ell)$ streams by (we need to know to which edge the bit $w_\ell$
corresponds), then the algorithm maps $\ket{i}\mapsto (-1)^{w_\ell}\ket{i}$ where $i=\min(i_\ell,j_\ell)$. At
the end, with probability $2\alpha$ the algorithm is left with a state
$\frac{1}{\sqrt{2}}((-1)^{x_{i_\ell}\oplus w_\ell}\ket{i_\ell}+(-1)^{x_{j_\ell}}\ket{j_\ell})$ for some edge
$(i_\ell,j_\ell)\in M$. The algorithm can learn the function value $x_{i_\ell}\oplus x_{j_\ell}\oplus w_\ell$
from this by a final measurement.

\subsection{Application: limits on classical simulation of quantum one-way protocols}\label{ssecsimulations}

A final application is in the context of simulating one-way quantum communication
protocols by one-way classical protocols. As noted by Aaronson~\cite[Section~5]{aaronson:qlearnability},
our Theorem~\ref{thseparationforalphapm} implies that his general simulation of bounded-error one-way
quantum protocols by deterministic one-way protocols
$$
D^1(f)=O(m Q^1(f)\log Q^1(f)),
$$
is tight up to a polylogarithmic factor. Here $m$ is the length of Bob's input.
This simulation works for any partial Boolean function $f$.  
Taking $f$ to be our \alphapm\ for $\alpha=1/4$,
one can show that $D^1(f)=\Theta(n)$, $m=\Theta(n\log n)$, and $Q^1(f)=O(\log n)$.

It also implies that his simulation of quantum bounded-error one-way protocols by classical bounded-error one-way protocols
$$
R^1(f)=O(m Q^1(f)),
$$
cannot be considerably improved. In particular, the product on the right cannot be replaced by the sum:
if we take $f=\alphapm$ with $\alpha=1/\sqrt{n}$, then by Theorem~\ref{thccseparation} we have
$R^1(f)\approx n^{3/4}$, $m\approx \sqrt{n}\log n$, and $Q^1(f)=O(\sqrt{n}\log n)$.

\section{The problem and its quantum and classical upper bounds}\label{secdefs}

We assume basic knowledge of quantum computation~\cite{nielsen&chuang:qc} and (quantum) communication
complexity~\cite{kushilevitz&nisan:cc,wolf:qccsurvey}.

Before giving the definition of our variant of the Boolean Hidden Matching Problem, we fix some
notation. Part of Bob's input will be a sequence $M$ of $\alpha n$ disjoint edges
$(i_1,j_1),\ldots,(i_{\alpha n},j_{\alpha n})$ over $[n]$, which we call an $\alpha$-matching.
We use ${\cal M}_{\alpha n}$ to denote the set of all such matchings.
If $\alpha=1/2$ then the matching is \emph{perfect}, if $\alpha < 1/2$ then the matching is \emph{partial}. We can view $M$ as
an $\alpha n \times n$ matrix over $GF(2)$, where the $\ell$-th row has exactly two 1s, at positions $i_\ell$ and $j_\ell$.
Let $x\in \01^n$. Then the matrix-vector product $Mx$ is an $\alpha n$-bit
string $z=z_1,\ldots ,z_\ell,\ldots z_{\alpha n}$ where  $z_\ell=x_{i_\ell}\oplus x_{j_\ell}$.
Using this notation, we define the following \emph{$\alpha$-Partial Matching (\alphapm)}
problem, whose one-way communication complexity we will study.

\medskip

\noindent
{\bf Alice:} $x\in\01^n$\\
{\bf Bob:} an $\alpha$-matching $M$ and a string $w\in \01^{\alpha n}$\\
{\bf Promise on the input:} there is a bit $b$ such that $w=Mx \oplus b^{\alpha n}$ (equivalently, $w=z$ or $w=\overline{z}$)\\
{\bf Function value:} $b$

\medskip

Actually, most of our analysis will not be concerned with Bob's second input $w$. Rather, we will show that
given only a short message about $x$, Bob will know essentially nothing about $z=Mx$. Note that to compute
$b$, it suffices that Bob learns one bit $z_\ell$ of the string $z$, since $b=z_\ell\oplus w_\ell$. We will
first give quantum and classical upper bounds on the message length needed for this.

\paragraph{\bf Quantum upper bound:}
Suppose Alice sends a uniform superposition of her bits to Bob:
$$
\ket{\psi}=\frac{1}{\sqrt{n}}\sum_{i=1}^n (-1)^{x_i}\ket{i}.
$$
Bob completes his $\alpha n$ edges to a perfect matching in an arbitrary way, and measures with the corresponding set of $n/2$
2-dimensional projectors. With probability $2\alpha$ he will get one of the edges $(i_\ell,j_\ell)$ of
his input $M$. The state then collapses to
$$
\frac{1}{\sqrt{2}}\left( (-1)^{x_{i_\ell}}\ket{i_\ell}+(-1)^{x_{j_\ell}}\ket{j_\ell}\right),
$$
from which Bob can obtain the bit $z_\ell=x_{i_\ell}\oplus x_{j_\ell}$ by measuring in the corresponding $\ket{\pm}$-basis.
Note that this protocol has so-called ``zero-sided error'': Bob knows when he didn't learn any bit $z_\ell$.
If Bob is given $O(k /\alpha)$ copies of $\ket{\psi}$, then with high probability
(at least while $k \ll \alpha n$) he can learn $k$ distinct bits of $z$.

{\bf Remark.} This protocol can be modified to a protocol in the simultaneous message passing model in a
standard way, first suggested by Buhrman (see \cite{gkrw:identification}). Alice and Bob share the maximally
entangled state $\frac{1}{\sqrt{n}}\sum_i \ket{i,i}$. Alice implements the transformation $\ket{i}
\rightarrow (-1)^{x_i}\ket{i}$ on her half. Bob performs the measurement with his projectors on his half. If
he gets one of the edges of his input, he sends the resulting $(i_\ell,j_\ell)$ and $w_\ell$ to the referee.
Now Alice and Bob perform a Hadamard transform on their halves, measure and send the result to the referee,
who has enough information to reconstruct $z_\ell$.

\paragraph{\bf Classical upper bound:}
We sketch an $O(\sqrt{n/\alpha})$ \emph{classical} upper bound. 
Suppose Alice uniformly picks a subset of $d\approx\sqrt{n/\alpha}$ bits of $x$ to send to Bob.
By the Birthday Paradox, with high probability Bob will have both endpoints of at least one of
his $\alpha n$ edges and so he can compute a bit of $z$ (and hence the function value $b$) with good probability.
In this protocol Alice would need to send about $d\log n$ bits to Bob, since
she needs to describe the $d$ indices as well as their bitvalues. However, by Newman's
Theorem~\cite{newman:random}, Alice can actually restrict her random choice to picking one out of $O(n)$
possible $d$-bit subsets, instead of one out of all ${n\choose d}$ possible subsets. Hence $d+O(\log n)$
bits of communication suffice. This matches our lower bound up to constant factors.


\section{Main proof}\label{secmainproof}

In this section we prove our main technical result (Theorem~\ref{thseparationforalphapm}), 
which shows that Bob knows hardly anything about the string $z=Mx$ unless Alice sends him a long message.

\subsection{Preliminaries}

We begin by providing a few standard definitions from Fourier analysis on the Boolean cube.
For functions $f,g:\01^n \rightarrow \mathbb{R}$ we define their inner product and $\ell_2$-norm by
\[
\inprod{f}{g}=\frac{1}{2^n}\sum_{x\in \01^n} f(x)g(x)\;\;\;,\;\;\;
\norm{f}_2^2 = \inprod{f}{f} = \frac{1}{2^n}\sum_{x\in\01^n} |f(x)|^2.
\]
The Fourier transform of $f$ is a function $\widehat{f}:\01^n\rightarrow\mathbb{R}$ defined by
\[
\widehat{f}(s) = \inprod{f}{\chi_s} = \frac{1}{2^n}\sum_{y\in\01^n} f(y)\chi_s(y),
\]
where $\chi_s:\01^n\rightarrow \mathbb{R}$ is the character $\chi_s(y)=(-1)^{y\cdot s}$ with ``$\cdot$'' being
the scalar product; $\widehat{f}(s)$ is the Fourier coefficient of $f$ corresponding to $s$.
We have the following relation between $f$ and $\widehat{f}$:
\[
f = \sum_{s\in\01^n} \widehat{f}(s)\chi_s.
\]
We will use two tools in our analysis, Parseval's identity and the KKL lemma.

\begin{lemma}[Parseval]\label{lem:parseval}
For every function $f:\01^n \rightarrow \mathbb{R}$ we have
$\displaystyle\norm{f}_2^2 = \sum_{s\in\01^n} \widehat{f}(s)^2.$
\end{lemma}

Note in particular that if $f$ is an arbitrary probability distribution on $\01^n$ and $U$ is the uniform
distribution on $\01^n$, then $\widehat{f}(0^n)=\widehat{U}(0^n)=1/2^n$ and $\widehat{U}(s)=0$ for nonzero $s$, hence
\begin{equation}\label{eqdistancefromuniform}
\norm{f-U}_2^2=\sum_{s\in\01^n} (\widehat{f}(s)-\widehat{U}(s))^2=\sum_{s\in\01^n\backslash\{0^n\}}\widehat{f}(s)^2.
\end{equation}

\begin{lemma}[\cite{kkl:influence}]\label{lem:kkl}
Let $f$ be a function $f:\01^n\rightarrow\{-1,0,1\}$. Let $A=\{x\mid f(x)\neq 0\}$,
and let $|s|$ denote the Hamming weight of $s\in\01^n$.
Then for every $\delta\in[0,1]$ we have
\[
\sum_{s\in\01^n} \delta^{|s|}\widehat{f}(s)^2 \leq \left(\frac{|A|}{2^n}\right)^{\frac{2}{1+\delta}}.
\]
\end{lemma}

We also need the following combinatorial lemma about uniformly chosen matchings.

\begin{lemma}\label{lemcountgoodmatchings}
Let $v\in\01^n$. If $|v|=k$ for even $k$, then
$$
\Pr_M[\exists\,s\in\01^{\alpha n}s.t.~M^Ts=v]=\frac{{\alpha n\choose k/2}}{{n\choose k}},
$$
where the probability is taken uniformly over all $\alpha$-matchings $M$.
\end{lemma}

\begin{proof}
We can assume without loss of generality that $v=1^k0^{n-k}$.
We will compute the fraction of matchings $M$ for which there exists such an $s$.
The total number of matchings $M$ of $\alpha n$ edges is $n!/(2^{\alpha n}(\alpha n)!(n-2\alpha n)!)$.
This can be seen as follows: pick a permutation of $n$, view the first $\alpha n$ pairs as $\alpha n$ edges,
and ignore the ordering within each edge, the ordering of the $\alpha n$ edges,
and the ordering of the last $n-2\alpha n$ vertices.
Note that $\exists~s~s.t.~M^Ts=v$ iff $M$ has exactly $k/2$ edges in $[k]$ and $\alpha n-k/2$ edges
in $[n]\backslash [k]$. The number of ways to pick $k/2$ edges in $[k]$ (i.e.~a perfect matching) is
$k!/(2^{k/2}(k/2)!)$. The number of ways to pick $\alpha n-k/2$ edges in $[n]-[k]$ is
$(n-k)!/(2^{\alpha n-k/2}(\alpha n-k/2)!(n-2\alpha n)!)$.
Hence the probability in the lemma equals
$$
\frac{k!/(2^{k/2}(k/2)!)~\cdot~(n-k)!/(2^{\alpha n-k/2}(\alpha n-k/2)!(n-2\alpha n)!)}{n!/(2^{\alpha n}(\alpha n)!(n-2\alpha n)!)}
=
\frac{{\alpha n\choose k/2}}{{n\choose k}}.
$$
\end{proof}

This probability is exponentially small in $k$ if $\alpha<1/2$, but it equals 1 if $\alpha=1/2$ and $v=1^n$.

\paragraph{\bf Total variation distance:}
For probability distributions $p$ and $q$ on the same finite set $S$, let
\begin{equation}\label{eqdeftvd}
\norm{p-q}_{tvd}=\sum_{i\in S}|p(i)-q(i)|
\end{equation}
denote their total variation distance.  This distance is 0 if $p=q$, it is 2 if $p$ and $q$ have support 
on disjoint sets, and between 0 and 2 otherwise.
Suppose we want to distinguish $p$ from~$q$: given only one sample 
we want to decide whether this sample came from $p$ or from $q$.
It is well known that the best success probability with which we can solve this task
is $1/2+\norm{p-q}_{tvd}/4$, so the total variation distance determines completely how well we can 
distinguish $p$ and $q$.

\subsection{The proof of Theorem~\ref{thseparationforalphapm}}

In order to prove Theorem~\ref{thseparationforalphapm},
consider any set $A\subseteq\01^n$ with $|A|\geq 2^{n-c}$ and let
$f:\01^n\rightarrow\01$ be its characteristic function (i.e.~$f(x)=1$ iff $x\in A$).
Let $\eps>0$, $\alpha\in(0,1/4]$, and $1\leq c\leq\gamma\eps\sqrt{n/\alpha}$ 
for some $\gamma$ to be determined later.

With $x$ uniformly distributed over $A$, we can write down Bob's induced distribution on $z$ as
$$
p_M(z)=\frac{|\{x\in A\mid Mx=z\}|}{|A|}.
$$
We want to show that $p_M$ is close to uniform, for most $M$.
By Eq.~(\ref{eqdistancefromuniform}), we can achieve this by bounding the Fourier coefficients of $p_M$.
These are closely related to the Fourier coefficients of $f$:
\begin{eqnarray}
\widehat{p_M}(s) &= & \frac{1}{2^{\alpha n}}\sum_{z\in\01^{\alpha n}} p_M(z)(-1)^{z\cdot s}\nonumber \\
 &=& \frac{1}{|A|2^{\alpha n}}\left(|\{x\in A \mid (Mx)\cdot s =0\}| - |\{x\in A \mid (Mx) \cdot s=1\}|\right)\nonumber \\
 &=& \frac{1}{|A|2^{\alpha n}}\left(|\{x\in A \mid x\cdot (M^Ts) =0\}| - |\{x\in A \mid x\cdot (M^Ts)=1\}|\right)\nonumber \\
 &=&\frac{1}{|A|2^{\alpha n}}\sum_{x \in \01^n} f(x)(-1)^{x\cdot (M^T s)}\nonumber \\
 &=& \frac{2^n}{|A|2^{\alpha n}}\cdot\widehat{f}(M^Ts).\label{eqfouriercoeff}
\end{eqnarray}
Note that the Hamming weight of $v=M^Ts\in\01^n$ is twice the Hamming weight of $s\in\01^{\alpha n}$.

Using KKL, we get the following bound on the level sets of the Fourier transform of~$f$:

\begin{lemma}\label{lemlevelsetsf}
For every $k\in\{1,\ldots,4c\}$ we have
$\displaystyle
\frac{2^{2n}}{|A|^2}\sum_{v:|v|=k}\widehat{f}(v)^2 \leq \left(\frac{4\sqrt{2}c}{k}\right)^k.$
\end{lemma}

\begin{proof}
By the KKL inequality (Lemma~\ref{lem:kkl}), for every $\delta\in[0,1]$ we have
\begin{equation*}
\frac{2^{2n}}{|A|^2}\sum_{v:|v|=k}\widehat{f}(v)^2  \leq
\frac{2^{2n}}{|A|^2}\frac{1}{\delta^k}\left(\frac{|A|}{2^n}\right)^{2/(1+\delta)}
 = \frac{1}{\delta^k}\left(\frac{2^n}{|A|}\right)^{2\delta/(1+\delta)}
 \leq \frac{1}{\delta^k}\left(\frac{2^n}{|A|}\right)^{2\delta}
 \leq \frac{2^{2\delta c}}{\delta^k}.
\end{equation*}
Plugging in $\delta=k/4c$ (which is in $[0,1]$ by our assumption on the value of $k$) gives the lemma.
\end{proof}

We bound the expected \emph{squared} total variation distance between $p_M$ and $U$ as follows:
\begin{align*}
\Exp_M[\norm{p_M-U}_{tvd}^2] & \leq 2^{2\alpha n}\,\Exp_M\left[\norm{p_M-U}_2^2\right]\\
 & = 2^{2\alpha n}\,\Exp_M\left[\sum_{s\in\01^{\alpha n}\backslash\{0^{\alpha n}\}}\widehat{p_M}(s)^2\right]\\
 & = \frac{2^{2n}}{|A|^2}\Exp_M\left[\sum_{s\in\01^{\alpha n}\backslash\{0^{\alpha n}\}}\widehat{f}(M^Ts)^2\right]\\
\intertext{where we used, respectively, the Cauchy-Schwarz inequality, Eq.~(\ref{eqdistancefromuniform}),
and Eq.~(\ref{eqfouriercoeff}).
Note that for each $v\in\01^n$, there is at most one $s\in\01^{\alpha n}$ for which $M^Ts=v$ (and the only $s$ that makes $M^Ts=0^n$, is $s=0^{\alpha n}$).
This allows us to change the expectation over $M$ into a probability and use Lemma~\ref{lemcountgoodmatchings}:}
 & = \frac{2^{2n}}{|A|^2}\Exp_M\left[\sum_{v\in\01^n\backslash\{0^n\}}|\{s\in\01^{\alpha n}\mid M^Ts=v\}|\cdot\widehat{f}(v)^2\right]\\
 & = \frac{2^{2n}}{|A|^2}\sum_{v\in\01^n\backslash\{0^n\}}\Pr_M\left[\exists\,s\in\01^{\alpha n}s.t.~M^Ts=v\right]\cdot\widehat{f}(v)^2\\
 & = \frac{2^{2n}}{|A|^2}\sum_{{\rm even}\, k=2}^{2\alpha n}\frac{{\alpha n\choose k/2}}{{n\choose k}}\sum_{v:|v|=k}\widehat{f}(v)^2.
\end{align*}
We first upper bound the part of this sum with $k<4c$.
Applying Lemma~\ref{lemlevelsetsf} for each $k$, using the standard estimates $(n/k)^k \leq {n\choose k}\leq (en/k)^k$,
and our upper bound $c\leq \gamma\eps\sqrt{n/\alpha}$, we get:
$$
\frac{2^{2n}}{|A|^2}\sum_{{\rm even}\, k=2}^{4c-2}\frac{{\alpha n\choose k/2}}{{n\choose k}}\sum_{v:|v|=k}\widehat{f}(v)^2
 \leq \sum_{{\rm even}\, k=2}^{4c-2}\frac{(2e\alpha n/k)^{k/2}}{(n/k)^k}\left(\frac{4\sqrt{2}c}{k}\right)^k
 \leq \sum_{{\rm even}\, k=2}^{4c-2}\left(\frac{64e\gamma^2\eps^2}{k}\right)^{k/2}.
$$
Picking $\gamma$ a sufficiently small constant, this is at most $\eps^2/2$ (note that the sum starts at $k=2$).

In order to bound the part of the sum with $k\geq 4c$, note that the function $g(k):={\alpha n\choose k/2}/{n\choose k}$
is decreasing for the range of even $k$ up to $2\alpha n$ (which is $\leq n/2$ because $\alpha\leq 1/4$):
$$
\frac{g(k-2)}{g(k)}=\frac{{\alpha n\choose k/2-1}/{n\choose k-2}}{{\alpha n\choose k/2}/{n\choose k}}
=\frac{(n-k+2)(n-k+1)k/2}{(\alpha n-k/2+1)(k-1)k}
=\frac{(n-k+2)(n-k+1)}{(2\alpha n-k+2)(k-1)}
\geq \frac{n-k+1}{k-1}\geq 1.
$$
We also have
$
\displaystyle\sum_{v\in\01^n}\widehat{f}(v)^2=\frac{|A|}{2^n}
$ 
by Parseval (Lemma~\ref{lem:parseval}), and $\displaystyle\frac{2^n}{|A|}\leq 2^c$ by assumption.
Hence
$$
\frac{2^{2n}}{|A|^2}\sum_{{\rm even}\, k=4c}^{2\alpha n}g(k)\sum_{v:|v|=k}\widehat{f}(v)^2
 \leq 2^c g(4c)
 \leq \left(\frac{8\sqrt{2}e\alpha c}{n}\right)^{2c}
 \leq \left(8\sqrt{2}e\gamma\eps\sqrt{\frac{\alpha}{n}}\right)^{2c}
 \leq \eps^2/2,
$$
where in the last step we used $\alpha/n\leq 1$ and $c\geq 1$, and 
picked $\gamma$ a sufficiently small constant.

Hence we have shown $\Exp_M[\norm{p_M-U}_{tvd}^2]\leq\eps^2$.
By Jensen's inequality we have 
$$
\Exp_M[\norm{p_M-U}_{tvd}]\leq \sqrt{\Exp_M[\norm{p_M-U}_{tvd}^2]}\leq\eps.
$$
This concludes the proof of

\bigskip

\noindent
{\bf Theorem~\ref{thseparationforalphapm}.}
{\it
Let $x$ be uniformly distributed over a set $A\subseteq\01^n$ of size $|A|\geq 2^{n-c}$ for some $c\geq 1$, 
and let $M$ be uniformly distributed over the set ${\cal M}_{\alpha n}$ of all $\alpha$-matchings, 
for some $\alpha\in(0,1/4]$.
There exists a universal constant $\gamma>0$ (independent of $n$, $c$, and $\alpha$), such that for all $\eps>0$:
if $c\leq \gamma\eps\sqrt{n/\alpha}$ then}
$$
\Exp_M\left[\norm{p_M - U}_{tvd}\right]\leq\eps.
$$

\medskip

The $\eps^2$ upper bound on $\Exp_M[\norm{p_M-U}^2_{tvd}]$ is essentially tight. 
This can be seen in the communication setting as follows.
With probability $\Omega(\eps^2)$ over the choice of $M$, 
at least one edge of $M$ will have both endpoints in
the first $c=\eps\sqrt{n/\alpha}$ bits. Then if Alice just sends the first $c$ bits of $x$ to Bob,
she gives him a bit of $z$. This makes $\norm{p_M-U}_{tvd}$ at least 1,
hence $\Exp_M[\norm{p_M-U}^2_{tvd}]=\Omega(\eps^2)$.

\subsection{The proof of Theorem~\ref{thccseparation}}

Our Theorem~\ref{thccseparation}, stated in the introduction, easily follows from Theorem~\ref{thseparationforalphapm}.
By the Yao principle \cite{yao:unified}, it suffices to analyze \emph{deterministic}
protocols under some ``hard'' input distribution. Our input distribution will be uniform over
$x\in\01^n$ and $M\in{\cal M}_{\alpha n}$. The inputs $x$ and $M$ together determine the $\alpha n$-bit
string $z=Mx$. To complete the input distribution, with probability $1/2$ we set $w=z$ 
and with probability $1/2$ we set $w$ to $z$'s complement $\overline{z}$.

Fix $\eps>0$ to a small constant, say $1/1000$. Let $c=\gamma\eps\sqrt{n/\alpha}$,
and consider any classical deterministic protocol that communicates at most $C=c-\log(1/\eps)$ bits.
This protocol partitions the set of $2^n$ $x$'s into $2^C$ sets $A_1,\ldots,A_{2^C}$, one for each possible message.
On average, these sets have size $2^{n-C}$. Moreover, by a simple counting argument,
at most a $2^{-\ell}$-fraction of all $x\in\01^n$ can sit in sets of size $\leq 2^{n-C-\ell}$.
Hence with probability at least $1-\eps$, the message that Alice sends corresponds to a set 
$A\subseteq\01^n$ of size at least $2^{n-C-\log(1/\eps)}=2^{n-c}$. 
In that case, by Theorem~\ref{thseparationforalphapm} and Markov's inequality, 
for at least a $(1-\sqrt{\eps})$-fraction of all $M$, the random variable $Z=MX$ 
(with $X$ uniformly distributed over $A$) is $\sqrt{\eps}$-close to the uniform distribution $U$. 
Given $w$, Bob needs to decide whether $w=Z$ or $w=\overline{Z}$. 
In other words, he is given one sample $w$, and needs to decide whether it came from 
distribution $Z$ or $\overline{Z}$. As we mentioned after Eq.~(\ref{eqdeftvd}), 
he can only do this if the distributions of $Z$ and $\overline{Z}$ have large total variation distance.
But by the triangle inequality
$$
\norm{Z-\overline{Z}}_{tvd}\leq\norm{Z-U}_{tvd}+\norm{\overline{Z}-U}_{tvd}=2\norm{Z-U}_{tvd}\leq 2\sqrt{\eps}.
$$
Hence Bob's advantage over randomly guessing the function value 
will be at most $\eps$ (for the unlikely event that $A$ is very small)
plus $\sqrt{\eps}$ (for the unlikely event that $M$ is such that $MX$ is more than $\sqrt{\eps}$ away from uniform) 
plus $\sqrt{\eps}/2$ (for the advantage over random guessing when $\norm{Z-U}\leq \sqrt{\eps}$).
To sum up: if the communication is much less than $\sqrt{n/\alpha}$ bits, then Bob cannot
decide the function value with probability significantly better than 1/2.

\section{The extractor-interpretation of our construction}\label{secextractor}

So far, we have proved that if the $n$-bit string $X$ is uniformly distributed over a set $A$ with $|A| \geq 2^{n-c}$
(i.e., a flat distribution on $A$), and $Y$ is uniformly distributed over all $\alpha$-matchings,
then $(Y,Z(X,Y))$ is close to uniform. In order to conclude the result about extractors mentioned in Section~\ref{ssecprivampl}, we need to prove the same result in the more general situation when $X$ has min-entropy greater than $n-c$ (instead of just being uniform on a set of size at least $2^{n-c}$). However, a result by Chor and Goldreich~\cite[Lemma~5]{CG} based on the fact that any distribution can be thought of as a convex combination of flat distributions, shows that the second statement follows from the first: flat distributions are the ``worst distributions'' for extractors.

\section{Conclusion}

In this paper we presented an extractor that is reasonably good when some small amount of \emph{classical} information
is known about the random source $X$ (technically: $H_{min}(X)\geq n-O(\sqrt{n/\alpha})$), but that fails
miserably if even a very small (logarithmic) amount of \emph{quantum} information is known about $X$.
We presented five applications of this:
\begin{enumerate}
\item An exponential quantum-classical separation for one-way communication complexity.
\item A classically-secure privacy amplification scheme that is insecure against a quantum adversary.
\item A key-expansion scheme that is secure against memory-bounded classical adversaries,
but not against quantum adversaries.
\item An exponential quantum-classical separation in the streaming model of computation.
\item The near-optimality of Aaronson's classical simulations of quantum one-way protocols.
\end{enumerate}
These applications all have the same flavor: they give examples where \emph{quantum memory is much more
powerful than classical memory}.
This contrasts for instance with the results about privacy amplification based on two-universal
hashing~\cite{kmr:powerofqmem,renner&koenig:privampl}, where quantum memory is not significantly
more powerful than classical memory.

\subsubsection*{Acknowledgments}
Many thanks to Oded Regev for his ideas on simplifying the two proofs in the conference
version of our paper to the current proof. Also many thanks to Renato Renner for explaining why
min-entropy determines the key length when doing privacy amplification
based on two-universal hashing (both in the classical and in the quantum case),
and him, Yevgeniy Dodis, Christian Schaffner, and Barbara Terhal for discussions about the bounded-storage model.
Thanks to Shengyu Zhang for bringing the streaming model to our attention, and to
Scott Aaronson for prodding us to make explicit the dependence on $\alpha$ in
Theorem~\ref{thseparationforalphapm} in light of the results mentioned in Section~\ref{ssecsimulations}.
Thanks to Guy Kindler and Jaikumar Radhakrishnan for various references that were needed for the earlier
proofs in our conference paper (but that are not needed anymore now).

\bibliographystyle{alpha}

\newcommand{\etalchar}[1]{$^{#1}$}

\end{document}